\documentstyle[aps,prb]{revtex}

\begin{document}

\input epsf.sty
\twocolumn[\hsize\textwidth\columnwidth\hsize\csname %
@twocolumnfalse\endcsname

\draft

\widetext

\title{Hole concentration dependence of the magnetic moment in 
superconducting and insulating La$_{2-x}$Sr$_{x}$CuO$_{4}$}

\author{S. Wakimoto$^{\star,\dagger}$ and R. J. Birgeneau$^{\dagger}$}
\address{Department of Physics and Center for Materials Science and 
Engineering, Massachusetts Institute of Technology, Cambridge, 
Massachusetts 02139}
\author{Y. S. Lee}
\address{National Institute of Standards and Technology, NCNR, 
Gaithersburg, Maryland 20889}
\author{G. Shirane}
\address{Physics Department, Brookhaven National Laboratory, Upton, New 
York 11973}

\date{\today}
\maketitle

\vspace{-0.1in}

\newpage
\begin{abstract}

Recent neutron scattering measurements on the La$_{2-x}$Sr$_{x}$CuO$_{4}$
system have revealed a drastic change of the incommensurate static spin correlations 
from diagonal in the insulating region to parallel in the
superconducting region.  
We report the doping dependence of the ordered magnetic moment 
%
%
for the hole concentration region $0.03 \leq x \leq 0.12$, 
focusing on the relationship
between the static magnetism and the superconductivity.
%
%
The elastic magnetic crosssection decreases monotonically with increasing 
$x$ for $0.03 \leq x \leq 0.07$.
We find that the 
ordered magnetic moment $\mu$ varies from $\sim0.18~\mu_{B}/$Cu ~$(x=0.03)$ to 
$\sim0.06 ~\mu_{B}/$Cu $(x=0.07)$.
No significant anomaly is observed at the insulator-superconductor boundary
$(x \sim 0.055)$.
The elastic magnetic cross section is enhanced in the vicinity of $x=0.12$ 
where resolution limited width peaks are observed in neutron scattering 
measurements
and where the apparent magnetic and superconducting transitions coincide.

\end{abstract}


\phantom{.}
]
\narrowtext

%
%
%

\section{Introduction}

%
The high-$T_{C}$ superconducting material La$_{2-x}$Sr$_{x}$CuO$_{4}$ 
exhibits a remarkable dependence of both its magnetism and its conductivity 
on the hole concentration $x$.~(Ref. 1)
%
%
In particular, the discovery of dynamic {\it incommensurate} (IC) spin correlations
in superconducting samples~\cite{Yoshizawa_88,Bob_89,S.W.Cheong_91}
stimulated investigations of the 
correlation between the microscopic magnetism and the high-$T_{C}$ superconductivity. 
Specifically, in superconducting samples inelastic IC magnetic peaks have 
been observed by  
neutron-scattering experiments at the IC positions with $\alpha \sim 45^{\circ}$ 
in polar coordinates as shown in the right inset of Fig. 1(a). 
We refer to these type of satellite peaks as ``parallel" incommensurate (PIC)
peaks.
%
%
Systematic neutron scattering experiments in the superconducting region 
by Yamada {\it et al.}~\cite{K.Yamada_98} revealed 
that the incommensurability $\delta$ of the PIC peaks~\cite{delta1} 
follows the linear relation $\delta=x$ for $0.06 \leq x \leq 0.12$.
%
%
Very sharp {\it elastic} IC magnetic peaks are also 
reported~\cite{T.Suzuki_98,Kimura_99} at the same 
PIC positions only in the vicinity of the special 1/8 hole concentration. 

%
On the other hand, Wakimoto {\it et al.}~\cite{waki_rapid,waki_full} found a 
new class of {\it elastic} IC magnetic peaks in the insulating $x=0.04$ and $0.05$ samples 
at the positions with $\alpha \sim 90^{\circ}$ as shown in the left inset of 
Fig. 1(a).
We refer to these peaks as ``diagonal" incommensurate (DIC) peaks.
%
%
Matsuda {\it et al.}~\cite{Matsuda_00} confirmed the existence of 
the static DIC phase down to Sr concentrations as low as $x=0.024$ as well as the linear relation 
$\delta=x$ in the insulating DIC phase.~\cite{delta2}
The detailed nature of the transition from the
insulator to the superconductor and concomitantly the DIC to the PIC magnetic phase 
has been also clarified recently.~\cite{Fujita_00}

%
It is clear from the phenomenological evidence discussed above that the IC 
nature of both the ``static" and ``dynamic" magnetic correlations relates 
closely to the superconductivity.
However, it is not yet fully understood what the intrinsic relation is
between the {\it static} moment and the superconductivity.  
%
%
Some previous neutron scattering studies have reported an
enhancement of the static component 
at the 1/8 hole concentration~\cite{T.Suzuki_98,Kimura_99,Tra_nature}
while other experimental techniques~\cite{other} 
have suggested a competitive relation.
On the other hand, 
coexistence of the superconductivity and static IC magnetic order has been 
reported
in stage-4 La$_{2}$CuO$_{4+\delta}$ by Lee {\it et al.}~\cite{Lee_99}
Furthermore, systematic muon spin resonance ($\mu$SR)~\cite{Ch.Niedermayer_98} 
and nuclear magnetic resonance~\cite{Chou_93PRL} measurements 
from the insulating region to the underdoped superconducting 
region have revealed that static (or quasi-static) 
magnetic order exists up to $x=0.10$ and that 
the spin-glass ordering temperature varies continuously across the 
insulator-superconductor boundary $x \sim 0.055$.
These results suggest that superconductivity and static magnetic 
order are at least compatible.

%
To understand the intrinsic relation between 
the superconductivity and the static magnetic order
we have carried out a quantitative comparison of the elastic 
magnetic neutron-scattering cross section over the concentration range 
$0.03 \leq x \leq 0.12$ taking 
into account the IC peak geometry. 
%
%
%
We deduce the ordered 
\linebreak
\begin{figure}
\centerline{\epsfxsize=2.5in\epsfbox{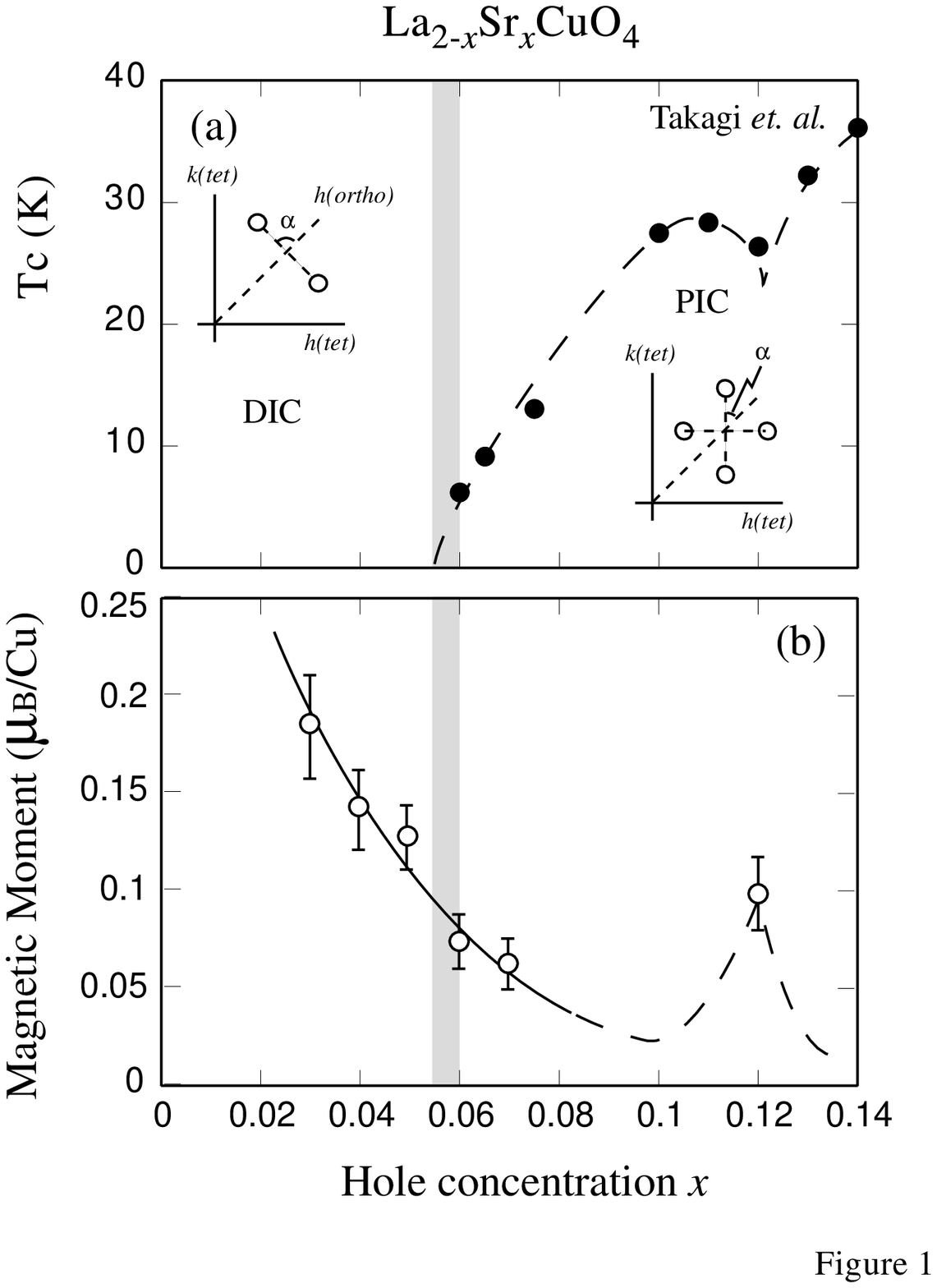}}
\caption{$x$ dependence of (a) superconducting transition temperature 
$T_{C}$ obtained from Ref. 23 and (b) magnetic moment.
The insets show the IC peak geometries in the reciprocal space.
Moment value for $x=0.12$ is reported in Ref. 22. 
Solid and dashed lines are guides to the eye.}
\end{figure}
\noindent
magnetic moment which 
contributes to the elastic cross-section, and 
find that the ratio between 
the statically ordered and 
dynamically fluctuating components 
changes systematically with $x$.
The Cu$^{2+}$ moment in undoped La$_{2}$CuO$_{4}$ is known 
to be $\sim 0.6~\mu_{\rm B}$/Cu.
Throughout this paper, we use Miller indices based on the orthorhombic 
$Bmab$ structure.

%
%

\section{Evaluation of magnetic moment for $x=0.05$}

\subsection{Fundamental formula}

Before we discuss the $x$-dependence of the magnetic moment, we first
present in this section the evaluation procedure for the magnetic moment in 
La$_{1.95}$Sr$_{0.05}$CuO$_{4}$.  
Once the moment in $x=0.05$ has been determined, the moments in the $x=0.03$ 
and $0.04$ samples can be estimated from the relative integrated intensity 
of the magnetic IC peaks as discussed in the next section.

From the definition of the magnetic structure factor, the calculated 
integrated intensity of the magnetic peaks can be described 
using the magnetic moment $\mu$ as below.

\begin{equation}
|F_{M}|^{2}_{cal} = p^{2}f^{2}({\rm {\bf Q}}) n^{2} 
\mu^{2} {\rm sin}^{2}\beta |F(h,k,l)|^2.
\label{eq_fm}
\end{equation}
In this formula, $pf({\rm {\bf Q}})$ is the neutron magnetic scattering length,
where $p=0.2696$~cm for $S=1/2$ spins. 
$f({\rm {\bf Q}})$ is the {\bf Q}-dependent magnetic form factor for the 
Cu$^{2+}$ spin that has been previously measured.~\cite{Shamoto}
The 
parameters
$n$, $\beta$ and $F(h,k,l)$ represent the number of spins in a magnetic unit 
cell, the angle of the Cu$^{2+}$ spins with 
respect to the scattering vector, 
and the magnetic structure factor, respectively.
The spin structure in the $x=0.05$ sample at low temperatures can be understood by 
a spin-glass cluster model~\cite{waki_full} in which each cluster has the undoped 
La$_{2}$CuO$_{4}$-type spin structure with random spin orientation in the 
plane.
With this structure, $n=4$ and $F(h,k,l)=1+e^{-\pi i(k+l)}-e^{-\pi i(h+k)}-e^{-\pi 
i(l+h)}$.  
Since the spin direction in each cluster is random, the factor 
${\rm sin}^{2}\beta$ in Eq.~(\ref{eq_fm}) should be modified to be 
$<{\rm sin}^{2}\beta>=1/2$, where $<\ >$ means an average over all of the clusters.

The relation between the calculated and observed 
integrated intensities is

\begin{equation}
\frac{|F_{M}|^{2}_{cal}}{|F_{N}|^{2}_{cal}} =
\frac{|F_{M}|^{2}_{obs}}{B |F_{N}|^{2}_{obs}} = A,
\label{eq_a}
\end{equation}
where $B$ is the extinction factor for the nuclear Bragg peak and the ratio $A$ is a constant
that can be determined experimentally.  
The indices $M$ and $N$ mean magnetic and nuclear scattering, respectively.
From Eqs.~(\ref{eq_fm}) and (\ref{eq_a}), the magnetic moment $\mu$ can be 
described as 

\begin{equation}
\mu^{2} = \frac{A |F_{N}|^{2}_{cal}}{p^{2} f^{2}({\rm {\bf Q}}) 
\cdot n^{2} <{\rm sin}^{2}\beta> |F(h,k,l)|^2}.
\label{eq_mu}
\end{equation}

\subsection{Integrated intensity}

%
In order to determine the parameter $A$, we made scans 
across the IC 
magnetic peaks and the (002) nuclear Bragg 
peak without changing the spectrometer 
configuration using the same $x=0.05$ crystal studied in Ref. 8.
The measurements were performed on the BT9 thermal-neutron triple-axis 
spectrometer installed at the NIST research reactor with the collimation 
sequence 20'-20'-S-20'-open and an incident neutron energy of 14.6~meV.
A Pyrolytic graphite filter was installed to eliminate contamination from higher-order 
neutrons.
%
%
For the evaluation of $|F_{N}|^{2}_{obs}$ and $|F_{N}|^{2}_{cal}$, we 
utilized the (002) peak that was found to have a small extinction factor 
$(B \sim 1)$ in a preliminary crystallographic experiment.
Actual scan profiles of the (002) peak and the IC peaks are shown in 
Figs. 2(a) and 2(b), respectively.
The scan trajectory for Fig. 2(b) is shown in the upper panel.
Both scans were made by changing only the sample rotation angle $\omega$.

Usually, integrated intensities for ``resolution-limited" peaks,
such as nuclear Bragg peaks, can be directly obtained from $\omega$-scans 
with 
$|F_{N}|^{2}_{obs} = R \int I(\omega)d\omega$.
Here $R$ is the Lorentz factor $1/\sin(2\theta)$ where $2\theta$ is the scattering 
angle and $I(\omega)$ is the measured intensity.
However, the IC magnetic peaks in $x=0.05$ have widths that are larger than 
the resolution.  
This is schematically illustrated in Fig. 3. 
If the peak width $\kappa_{a}$ is larger than the resolution 
width $W_{a}$, the integration along the one-dimensional trajectory shown 
by an arrow gives the integrated intensity 
\linebreak
\begin{figure}
\centerline{\epsfxsize=2.5in\epsfbox{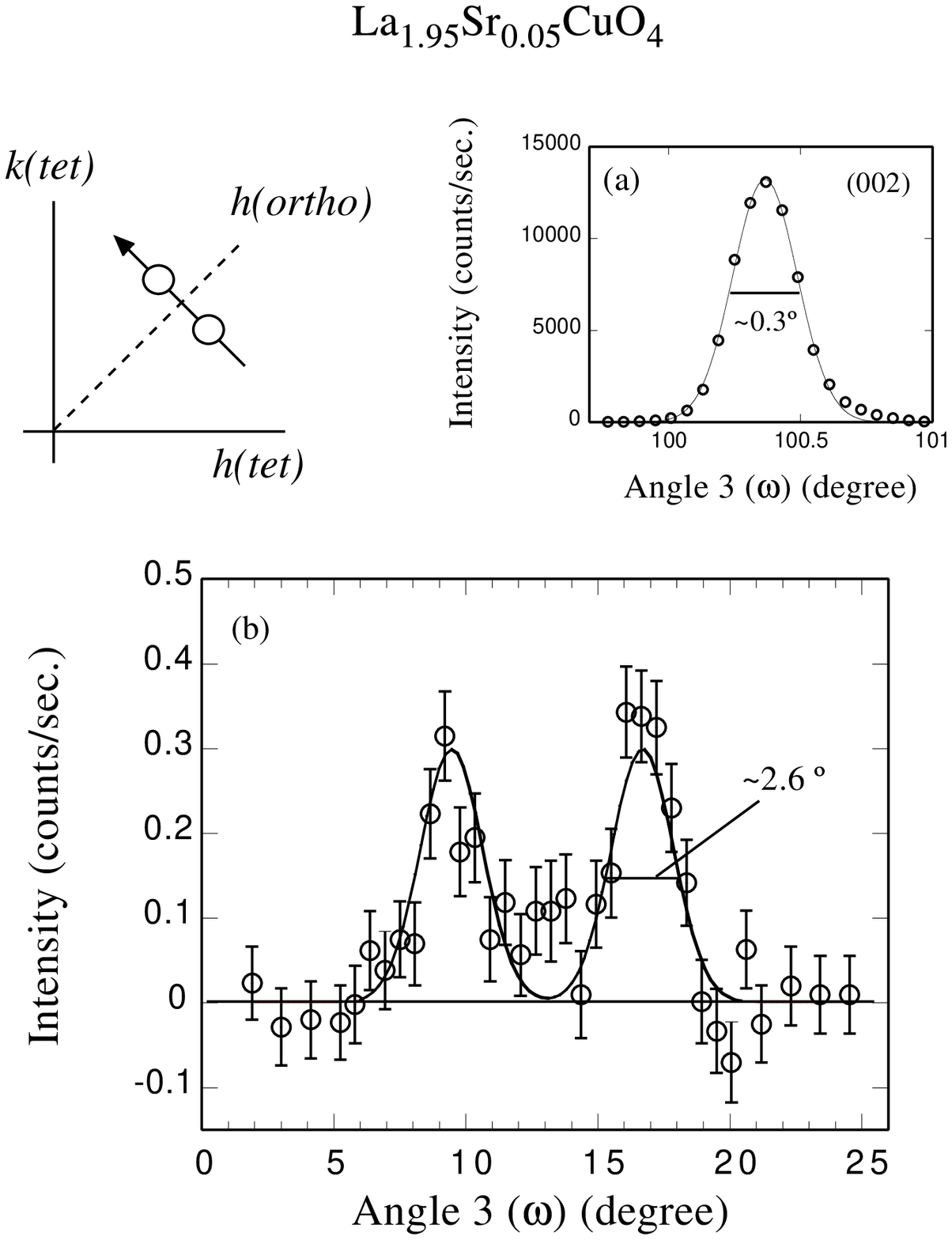}}
\caption{Peak profiles of (a) the (002) peak and (b) the IC elastic 
magnetic peaks for $x=0.05$.
Both are measured by changing sample-rotation angle $\omega$.
The scan profile for (b) is shown in the left upper panel.
Solid lines are fits by the Gaussian function.}
\end{figure}
\noindent
only of the A part of the peak cone, and
the intensities in the B parts will be missed.  

In our measurements of the IC peaks around (100), 
the $\kappa_{a}$ and $W_{a}$ values are 0.094 and 0.0165 \AA$^{-1}$, 
respectively, and $\kappa_{b}$ is 0.045 \AA$^{-1}$.
In this configuration, the actual in-plane integrated intensity for one IC peak should 
be described as $|F'|^{2}_{obs} = 1.6 R \int I(\omega)d\omega$.
The factor 1.6 is the volume ratio of the total peak cone and the A part in 
Fig. 3.

The same consideration must be taken into account for the peak width 
along the direction perpendicular to the scattering plane, that is, the $c^{*}$ 
direction.
As reported in Ref. 10, the $L$ dependence of the IC peak in the $x=0.05$ 
sample is very broad. Thence, we should utilize the intensity integrated over 
the Brillouin zone, that is, for $-1 \leq L \leq 1$.
From the vertical instrumental resolution ($\sim 0.13$~\AA$^{-1}$),
the correction factor for the peak spread along the $c^{*}$ direction 
is estimated to be $4 (\pm 0.5)$, which multiplies $|F'|^{2}_{obs}$.
%
%

To evaluate the parameter $A$, we utilized  
$|F_{M}|^{2}_{obs}=\sum |F'|^{2}_{obs}$, where the summation is made for 
every IC peak around the (100) position for all orthorhombic-twin domains.
With the procedure above and Eq.~(\ref{eq_mu}), we finally obtained 
$\mu \sim 0.13 (\pm 0.02)~\mu_{\rm B}/{\rm Cu}$ for $x=0.05$.

\section{Results and discussion}

%
The magnetic properties of the $x=0.03$ and $0.04$ samples are 
essentially the same as those of the $x=0.05$ system showing 
spin-glass behavior~\cite{waki_sg} and 
the DIC 
\linebreak
\begin{figure}
\centerline{\epsfxsize=2.2in\epsfbox{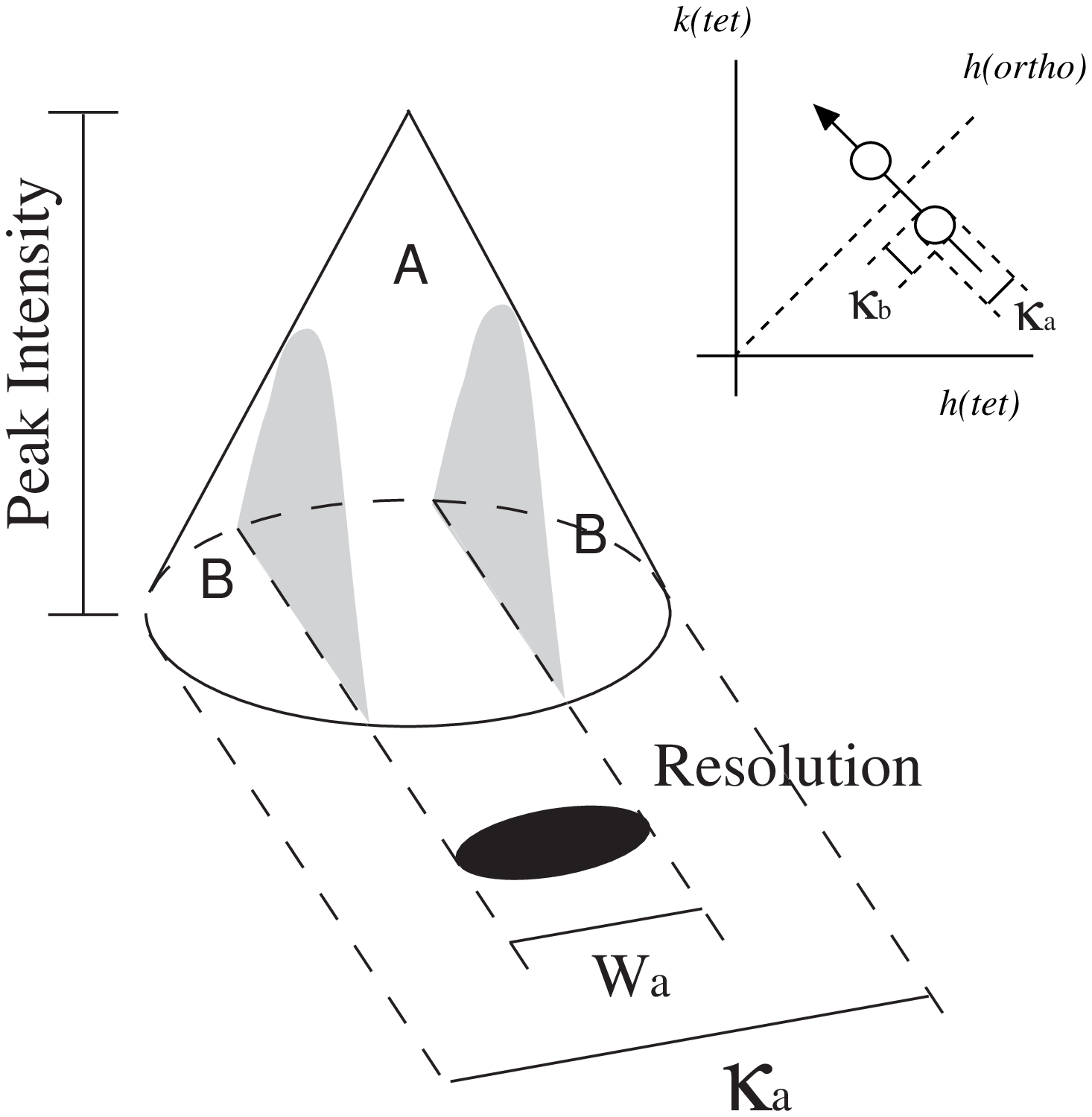}}
\caption{Schematic illustration showing the broad peak and instrumental 
resolution.  $\kappa_{a}$ and $\kappa_{b}$ correspond to the full width at 
the half maximum of the IC magnetic peak along the orthorhombic $a^{*}$ and 
$b^{*}$ axis.
$W_{a}$ is the instrumental resolution width along the $a^{*}$ axis.}
\end{figure}
\noindent
peaks.~\cite{waki_full,Matsuda_00}
These facts suggest the same magnetic structure in these compositions.
Therefore, the magnetic moments for $x=0.03$ and $0.04$ can be estimated 
by a direct comparison of $|F_{M}|^{2}_{obs}$ normalized by sample volume.
%
%
For $x=0.06$ and $0.07$, we also compared the normalized 
$|F_{M}|^{2}_{obs}$ from the elastic cross sections reported in 
Ref. 13 that demonstrated that the $x=0.06$ sample shows DIC and PIC 
components while the $x=0.07$ sample shows only PIC components.
The magnetic moments for these compositions are also evaluated from their 
normalized 
$|F_{M}|^{2}_{obs}$ calculated by summation of the integrated intensities of 
all the IC peaks around (100).

We note that the $L$ dependence of the IC peaks in the $x=0.024$ material
as reported by Matsuda {\it et al.}~\cite{Matsuda_00} is clearer 
than that in the $x=0.05$ sample.
This means that the correction factor for the peak spread along 
the $c^{*}$ direction should decrease with decreasing $x$.
Thence, we estimated the correction factors for $x=0.03$ and $0.04$ 
to be between 3 and 4, where the lower value 3 was calculated from the 
$L$ dependence of $x=0.024$.
The ambiguity of this factor is shown as error bars in Fig. 1(b).
For $x=0.06$ and $0.07$, the $L$-correction factor of $4 (\pm 0.5)$ was 
utilized. 


The resultant magnetic moments together with that for $x=0.12$ reported by Kimura 
{\it et al.}~\cite{Kimura_00} are summarized in Fig. 1(b).
Figure 1(a) shows the $x$ dependence of $T_{C}$ reported by Takagi {\it et al.}~\cite{Takagi}
First we discuss the $x$ dependence of the magnetic moment near the 
insulator-superconductor boundary $x \sim 0.055$.
The magnetic moment decreases monotonically across the boundary 
although we cannot exclude the possiblity of a small drop at the boundary. 
%
%
In a previous study in the lightly doped region, it was reported that the 
elastic magnetic cross section was constant for $0.03 \leq x \leq 0.05$ 
and suddenly decreased at the 
insulator-superconductor boundary.~\cite{waki_isspprc}
However, in that analysis, the integration of the cross section was made only 
along the one-dimensional scan so that the cross section outside the instrumental 
resolution was not properly taken into account.
Moreover, the correct IC peak geometry was not known at that time.

Other characteristics of the elastic IC peak have been reported~\cite{Fujita_00}
to be continuous across the insulator-superconductor boundary,
such as the peak width and the onset temperatures where the elastic IC peaks 
become observable.
Thus, the only dramatic change in the magnetic state at the boundary is the change of 
the IC modulation direction.
This collection of evidence suggests that it is 
the transition from the DIC to PIC magnetic state rather than any decrease 
of the static ordered component that strongly correlates with superconductivity.


In Fig. 1(b) we draw the dashed line as a guide to the eye so that the moment has a 
sharp maximum at $x=0.12$ since the sharp and intense IC elastic 
peaks have only been observed in the vicinity of $x=0.12$.~\cite{Kimura_99} 
%
%
%
It should be noted that the ordered phases in samples with $0.03 \leq x \leq 
0.07$ compared with that in $x=0.12$ are different.
The former has a small correlation length $\xi \sim 
20$~\AA, while the latter shows resolution-limited peaks that correspond 
to $\xi \geq 200$~\AA.
Such a long-range ordered state might affect the superconductivity 
differently from the quasistatic glassy state observed in $0.03 \leq x \leq 0.07$.
Further study of the intrinsic difference between these ordered states
is required to clarify the relation between the static magnetic order and the 
superconductivity. 

Finally, we briefly mention the recent results of $\mu$SR measurements
on the La$_{2-x}$Sr$_{x}$CuO$_{4}$ system by Uemura {\it et al.}~\cite{Uemura}
They reported that the ordered magnetic moment is almost constant at 
$\sim0.3~\mu_{\rm B}/$Cu in La$_{1.88}$Sr$_{0.12}$CuO$_{4}$ and  
stage-4 La$_{2}$CuO$_{4+\delta}$, however they infer that the volume fraction 
that contributes to the statically ordered signal may be 
different.~\cite{Assa}
We should note that the systematic change of the ordered moment in the 
lightly doped region as reported in this paper may be caused by the same 
feature,
that is, it is possible that only the volume fraction of the ordered phase 
instead of the ordered moment varies with $x$,
since we cannot uniquely determine the volume fraction from our neutron scattering 
measurements.

\section{Acknowledgments}

We thank Y. Endoh, M. Fujita, K. Hirota, H. Hiraka, M. A. Kastner, H. Kimura, 
S. H. Lee, M. Matsuda, T. Uemura, K. Yamada and A. Zeludev for valuable discussions. 
The present work was supported by the US-Japan Cooperative Research 
Program on Neutron Scattering.  
The work at MIT was supported by the NSF under Grant No.\ DMR0071256 
and by the MRSEC Program of the National Science Foundation under 
Award No.\ DMR98-08941.  
The work at Brookhaven National Laboratory was carried out under 
Contract No.\ DE-AC02-98CH10886, Division of Material Science, U.S. 
Department of Energy. 
The work at NIST is based upon activities supported by the National 
Science Foundation under Agreement No. DMR-9423101.

\vspace{5mm}
\noindent
$\star$ Also at Brookhaven National Laboratory, Upton, New York 11973

\noindent
$\dagger$ Present address: Department of Physics, University of Toronto, Toronto, Ontario, Canada M5S 1A7


\begin{references}

\bibitem{M.A.Kastner_98} M. A. Kastner, R. J. Birgeneau, G. Shirane, and Y. Endoh, 
Rev.\ Mod.\ Phys. {\bf 70}, 897 (1998).

\bibitem{Yoshizawa_88} H. Yoshizawa, S. Mitsuda, H. Kitazawa, and K. Katsumata,
J.\ Phys.\ Soc.\ Jpn. {\bf 57}, 3686 (1988).

\bibitem{Bob_89} R. J. Birgeneau, Y. Endoh, Y. Hidaka, K. Kakurai, M. A. Kastner, 
T. Murakami, G. Shirane, T. R. Thurston, and K. Yamada,
Phys.\ Rev.\ B {\bf 39}, 2868 (1989);
T. R. Thurston, R. J. Birgeneau, M. A. Kastner, N. W. Preyer, G. Shirane, Y. Fujii,
K. Yamada, Y. Endoh, K. Kakurai, M. Matsuda, Y. Hidaka, and T. Murakami,
{\it ibid}. {\bf 40}, 4585 (1989).

\bibitem{S.W.Cheong_91} S.-W. Cheong, G. Aeppli, T. E. Mason, H. A. Mook, 
S. M. Hayden, P. C. Canfield, Z. Fisk, K. N. Klausen, and J. L. Martinez, 
Phys.\ Rev.\ Lett. {\bf 67}, 1791 (1991).

\bibitem{K.Yamada_98} K. Yamada, C. H. Lee, K. Kurahashi, J. Wada, S. Wakimoto, 
S. Ueki, H. Kimura, Y. Endoh, S. Hosoya, G. Shirane, R. J. Birgeneau, M. Greven, 
M. A. Kastner, and Y. J. Kim, 
Phys.\ Rev.\ B {\bf 57}, 6165 (1998). 

\bibitem{delta1} $\delta$ can be defined by the distance between $(\pi, 
\pi)$ and the PIC peak positions
$(1/2 \pm \delta, 1/2)$ and $(1/2, 1/2 \pm \delta)$ in the tetragonal 
{\it I4/mmm} notation.

\bibitem{T.Suzuki_98} T. Suzuki, T. Goto, K. Chiba, T. Shinoda, T. Fukase, 
H. Kimura, K. Yamada, M. Ohashi, and Y. Yamaguchi, 
Phys.\ Rev.\ B {\bf 57}, 3229 (1998).

\bibitem{Kimura_99} H. Kimura, K. Hirota, H. Matsushita, K. Yamada, Y. Endoh,
S.-H. Lee, C. F. Majkrzak, R. Erwin, G. Shirane, M. Greven, Y. S. Lee, M. A. Kastner, 
and R. J. Birgeneau, 
Phys.\ Rev.\ B {\bf 59}, 6517 (1999); (private communication).

\bibitem{waki_rapid} S. Wakimoto, G. Shirane, Y. Endoh, K. Hirota, S. Ueki, K. Yamada, 
R. J. Birgeneau, M. A. Kastner, Y. S. Lee, P. M. Gehring, and S. H. Lee,
Phys.\ Rev.\ B {\bf 60}, R769 (1999).

\bibitem{waki_full} S. Wakimoto, R. J. Birgeneau, M. A. Kastner, Y. S. Lee,
R. Erwin, P. M. Gehring, S. H. Lee, M. Fujita, K. Yamada, Y. Endoh, K. 
Hirota, and G. Shirane, 
Phys.\ Rev.\ B {\bf 61}, 3699 (2000).

\bibitem{Matsuda_00} M. Matsuda, M. Fujita, K. Yamada, R. J. Birgeneau, M. 
A. Kastner, H. Hiraka, Y. Endoh, S. Wakimoto, and G. Shirane,
Phys.\ Rev.\ B. {\bf 62}, 9148 (2000).

\bibitem{delta2} $\delta$ of DIC peaks is defined by a distance between 
$(\pi, \pi)$ and the DIC peaks in the {\it tetragonal} reciprocal lattice unit.
Therefore, DIC peak positions can be described as  
$(1, \pm \sqrt2\delta)$ in the orthorhombic {\it Bmab} notation.

\bibitem{Fujita_00} M. Fujita, K. Yamada, H. Hiraka, S. H. Lee, P. M. 
Gehring, S. Wakimoto, G. Shirane,
cond-mat/0101320.

\bibitem{Tra_nature} J. M. Tranquada, B. J. Sternlieb, J. D. Axe, Y. 
Nakamura, S. Uchida,
Nature {\bf 375}, 561 (1995);
J. M. Tranquada, J. D. Axe, N. Ichikawa, Y. Nakamura, S. Uchida, and B. 
Nachumi,
Phys.\ Rev.\ B {\bf 54}, 7489 (1996).

\bibitem{other} For example, by muon spin rotation ($\mu$SR)
measurement in 
K. Kumagai, K. Kawano, I. Watanabe, K. Nishiyama, and K. Nagamine,
J. Supercond. {\bf 7}, 63 (1994).
By nuclear magnetic resonance (NMR) measurement in 
T. Goto, S. Kazama, K. Miyagawa, and T. Fukase, 
J. Phys. Soc. Jpn. {\bf 63}, 3494 (1994).

\bibitem{Lee_99} Y. S. Lee, R. J. Birgeneau, M. A. Kastner, Y. Endoh, S. Wakimoto, K. Yamada, 
R. W. Erwin, S. H. Lee, and G. Shirane,
Phys.\ Rev.\ B {\bf 60}, 3643 (1999).

\bibitem{Ch.Niedermayer_98} Ch. Niedermayer, C. Bernhard, T. Blasius, 
A. Golnik, A. Moodenbaugh, and J. I. Budnick, Phys.\ Rev.\ Lett. {\bf 80}, 
3843 (1998).

\bibitem{Chou_93PRL} F. C. Chou, F. Borsa, J. H. Cho, D. C. Johnston, A. 
Lascialfari, D. R. Torgeson, and J. Ziolo,
Phys.\ Rev.\ Lett. {\bf 71}, 2323 (1993).



\bibitem{Shamoto} S. Shamoto, M. Sato, J. M. Tranquada, B. J. Sternlieb, and 
G. Shirane,
Phys.\ Rev.\ B {\bf 48}, 13817 (1993).

\bibitem{waki_isspprc} S. Wakimoto, K. Yamada, S. Ueki, G. Shirane, Y. S. 
Lee, S. H. Lee, M. A. Kastner, K. Hirota, P. M. Gehring, Y. Endoh, and R. J. 
Birgeneau,
J.\ Phys.\ Chem.\ Solids {\bf 60}, 1079 (1999).

\bibitem{waki_sg} S. Wakimoto, S. Ueki, Y. Endoh, and K. Yamada,
Phys.\ Rev.\ B {\bf 62}, 3547 (2000).

\bibitem{Kimura_00} H. Kimura, H. Matsushita, K. Hirota, Y. Endoh, K. Yamada,
G, Shirane, Y. S. Lee, M. A. Kastner, R. J. Birgeneau,
Phys.\ Rev.\ B {\bf 61}, 14366 (2000); (private communication).

\bibitem{Takagi} H. Takagi, T. Ido, S. Ishibashi, M. Uota, and S. Uchida,
Phys.\ Rev.\ B {\bf 40}, 2254 (1989).

\bibitem{Uemura} T. Uemura (private communication);
T. Uemura, Y. S. Lee, and R. J. Birgeneau (unpublished).

\bibitem{Assa} $x$-independence of the magnetic moment has been also
suggested by numerical calculations: A. Auerbach (private communication);
M. Havilio and A. Auerbach, Phys.\ Rev.\ Lett. {\bf 83}, 4848 (1999).

\end{references}
\end{document}